\documentclass[12pt,notoc]{JHEP3}
\usepackage{epsfig}
\usepackage{amsmath}
\title{
Holographic Nuclei : \\
Supersymmetric Examples
}
\author{
Koji Hashimoto\\
Theoretical Physics Lab.,  
Nishina Center, 
RIKEN, Saitama 351-0198, Japan\\
E-mail: \email{koji@riken.jp}\\
}

\abstract
{
We provide a dual gravity description of a supersymmetric heavy
nucleus, following the idea of our previous paper arXiv/0809.3141. 
The supersymmetric nucleus consists of a merginal bound
state of $A$ baryons distributed over a ball in 3 dimensions.
In the gauge/string duality, the baryon in ${\cal N}=4$ super 
Yang-Mills (SYM) theory corresponds to 
a D5-brane wrapping $S^5$ of the $AdS_5\times S^5$ spacetime,
so the nucleus corresponds to a collection of $A$ D5-branes.
We take a large $A$ and a near horizon limits of a back-reacted geometry
generated by the wrapped $A$ D5-branes, where we find a gap in the
supergravity fluctuation spectrum. This spectrum 
is a gravity dual of giant resonances of heavy nuclei, in the
supersymmetric toy example of QCD.
}

\preprint{
{\normalsize RIKEN-TH-174}
}

\begin{document}

\section{Gauge/string duality and heavy nuclei}
\label{section1}

Heavy nuclei are bound states of large number of nucleons. Despite their
importance in physics, a fundamental understanding of it based on QCD is
still missing. This is partly because of complication of the many-body
system, where mass number $A$ of a nucleus is large. The
gauge/string duality \cite{Maldacena:1997re,Gubser:1998bc} (the AdS/CFT
correspondence) opened a new 
path to solve strongly coupled 
gauge theories such as large $N$ QCD, in which essentially a 
large number of
D-branes are necessary to validate the dual gravity description.
Baryons generically 
correspond to, in the gravity description of large $N$ QCD-like theories, 
D-branes wrapping
nontrivial cycles (called baryon vertices) \cite{Gross-Ooguri}.
In our previous paper \cite{Hashimoto:2008jq}, we noticed that, 
{\it we can take a large $A$ limit and a near horizon limit of the
wrapped D-branes for extracting the 
collective physics of $A$ baryons, i.e.~physics of a heavy nucleus.} 
This procedure is equivalent to
looking at a particular sub-sector of the holographic QCD. 
This idea was initiated in our previous paper \cite{Hashimoto:2008jq},
and
a schematic picture of the procedure is shown in Fig.~\ref{fig1}.

In this paper, we provide supersymmetric examples realizing this idea. 
In string theory, supersymmetric setups are more familiar, and our
example in this paper is for the ${\cal N}=4$ supersymmetric Yang-Mills
(SYM) theory for which the original gauge/string duality was
conjectured. We will find a near horizon geometry produced by a 
large number of baryon 
vertices ($A$ D5-branes wrapping the $S^5$ with $A \to\infty$). 
Fluctuation analysis of the background geometry provides, through the
gauge/string dictionary, a
spectrum of a collective motion of the baryons, {\it i.e.} giant
resonances of a heavy nucleus with mass number $A$.

\FIGURE[h]{
\includegraphics[width=9cm]{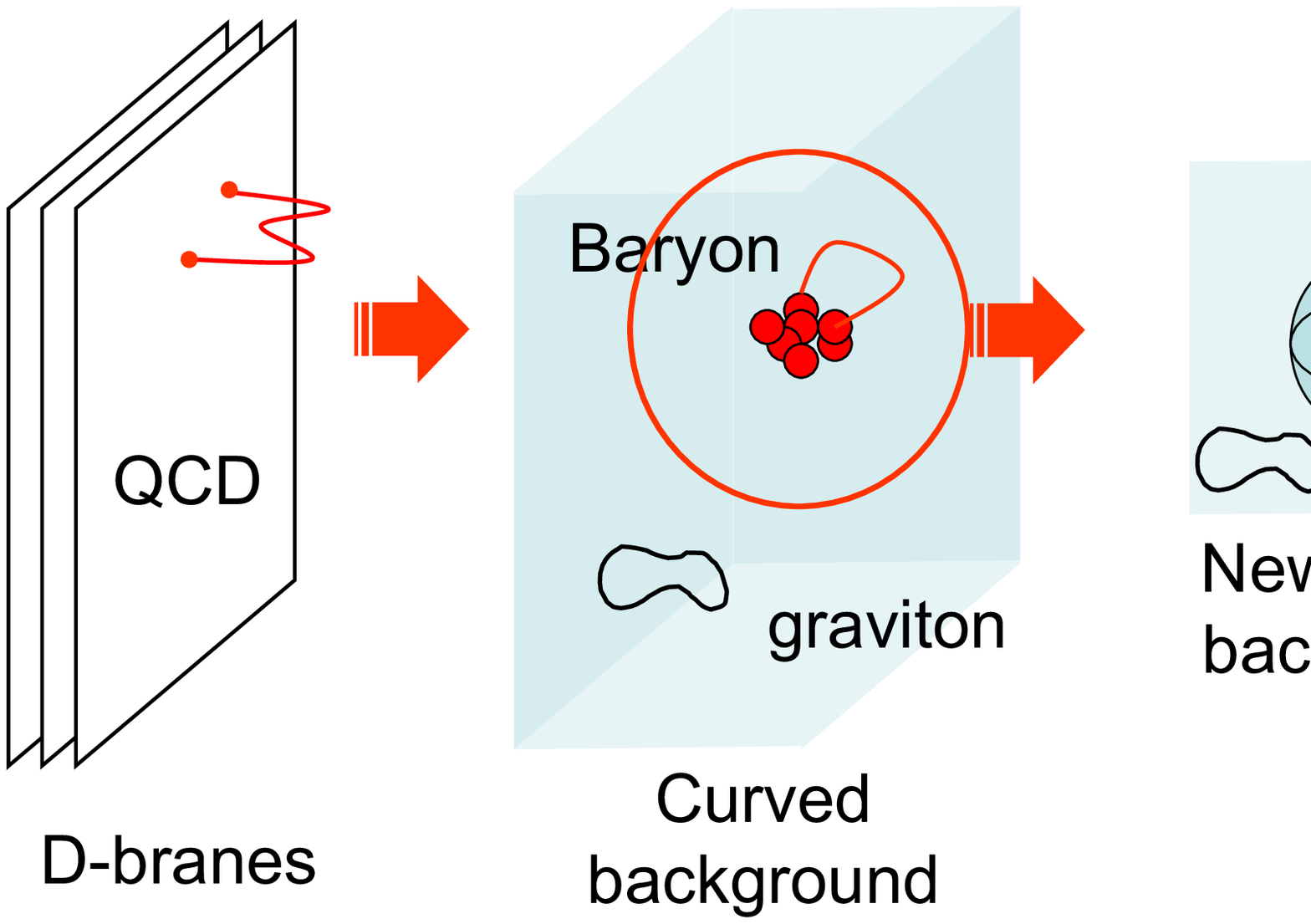}
\caption{How to obtain gravity backgrounds dual to the heavy nuclei. The
left arrow means taking a geometry dual to a large $N$ QCD. Then
one has ``baryon vertices'' (= particular D-branes wrapping cycles) in
the geometry. The right arrow shows the procedure studied in this paper.
We consider further a near horizon geometry of the back-reacted 
geometry produced by the baryon vertices. 
This corresponds to looking at a sub-sector of the whole holographic
QCD, which describes collective excitations on the heavy
nucleus.} 
\label{fig1}
}

\section{Geometry around baryon vertices}

Let us start by reviewing basic properties of the baryon vertex 
\cite{Gross-Ooguri} in the $AdS_5\times S^5$ geometry in type IIB 
superstring
theory. This geometry is
a gravity description of the ${\cal N}=4$ SYM in the large $N$ and the
large 'tHooft coupling limits, and the baryon corresponds to a D5-brane
wrapping the $S^5$ and localized in the radial direction of the $AdS_5$
as $r=r_0$, called ``baryon vertex''. 

\FIGURE[t]{
\includegraphics[width=5cm]{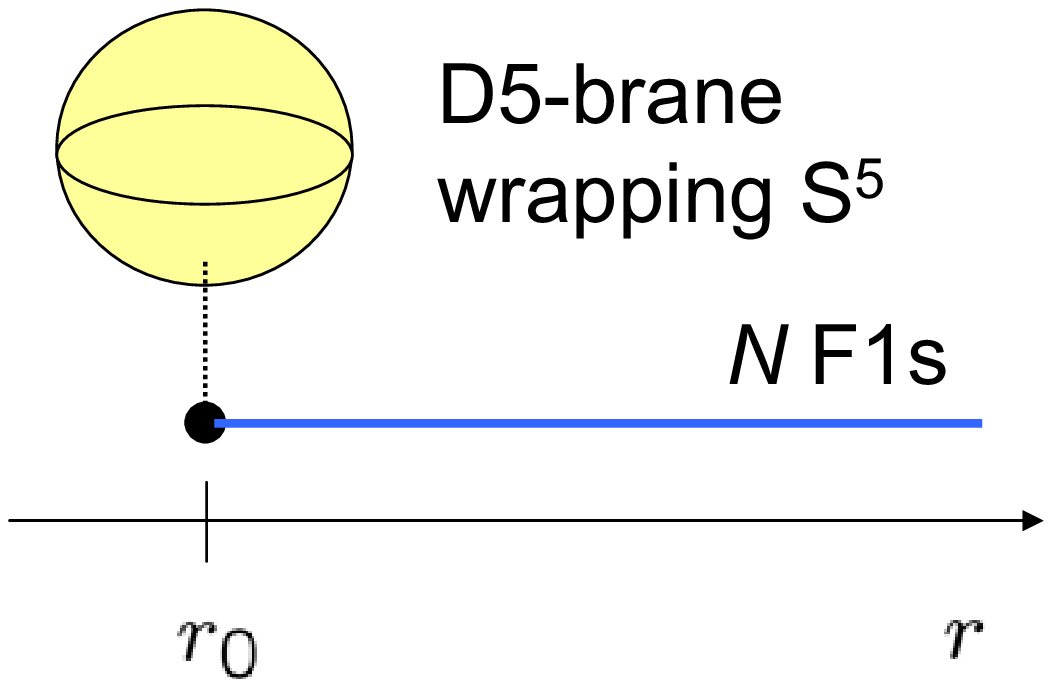}
\vspace{-1em}
\caption{The baryon vertex.}
\label{fig2}
}
The reason why the D5-brane corresponds to a baryon in the
boundary field theory is as follows.
The effective action of the D5-brane has a
Chern-Simons term of the form
\begin{eqnarray}
 \int_{S^5\times R} C_4\wedge F \sim \int dC\wedge A_0dx^0,
\label{css5}
\end{eqnarray}
meaning that the background Ramond-Ramond 5-form flux 
$dC_4 \propto N d\Omega^{(S^5)}$ in the geometry 
generates $N$ electric charges on the
D5-brane. The charge provides an electric flux going out
from the D5-brane, which is $N$
fundamental strings (see Fig.~\ref{fig2}),
which are interpreted as (infinitely massive) $N$ quarks in the SYM.

 One can solve the BPS equations of motion of the
D5-brane action \cite{Imamura:1998gk}, and can see that indeed the
D5-brane configuration is supersymmetric. The fundamental strings
attached to it is represented by a spike configuration
\cite{Callan:1997kz}. 
According to the scale-radius
correspondence of the gauge/string duality,
the size $z_0$ of the
baryon measured in the boundary SYM theory is related to $r_0$, the
position of the D5-brane in the $AdS_5$. 
Borrowing a similar 
interpretation of a SYM instanton size in terms of the bulk location
of a D(-1)-brane (for example see \cite{Balasubramanian:1998de}), we
have  
\begin{eqnarray}
 z_0 = \alpha'\sqrt{2\lambda}\; r_0^{-1}
\label{radius}
\end{eqnarray}
where $\lambda$ is the 'tHooft coupling of the SYM.
It was found \cite{Imamura:1998gk} 
that the total energy of the baryon
vertex does not depend on $r_0$, and 
is equal to that of $N$ fundamental strings
extending from the $AdS$ horizon to $AdS$ boundary: 
$E_{\rm D5 \; spike} = N E_{\rm F1}$. 

Now, let us turn to the nucleus. Any nucleus is a collection of $A$
nucleons, so it is just $A$ D5-branes accumulated. 
Therefore, the low energy theory on the nucleus is a non-Abelian
D5-brane effective theory, which is a
$U(A)$ SYM theory on spatial $S^5 \times$ time ${\bf R}$.

Since all of the $A$ baryons preserve the
supersymmetries, there is no force between the baryon vertices. This
means that we do not have any bound state forming the nucleus, due to the
supersymmetries. Rather to say, we may consider an arbitrary distribution
of baryon vertices. So, as an example, let us distribute 
the baryon vertices uniformly on an $S^3$ in the $AdS_5$ for
simplicity. Since each 
bulk location of the component baryon vertex corresponds to the size and
the location of the baryon in the boundary SYM, this spherical
distribution amounts to a ball-like distribution of baryons in the SYM
(see Fig.~\ref{fig3}). We chose this distribution by two reasons: First,
it mimics real nuclei, the simplest among which have the shape of a
ball. Second, with this spherical distribution of the baryon vertices,
it is easy to compute a fluctuation spectrum in the geometry
created by the baryon vertices.

\FIGURE[t]{
\includegraphics[width=5cm]{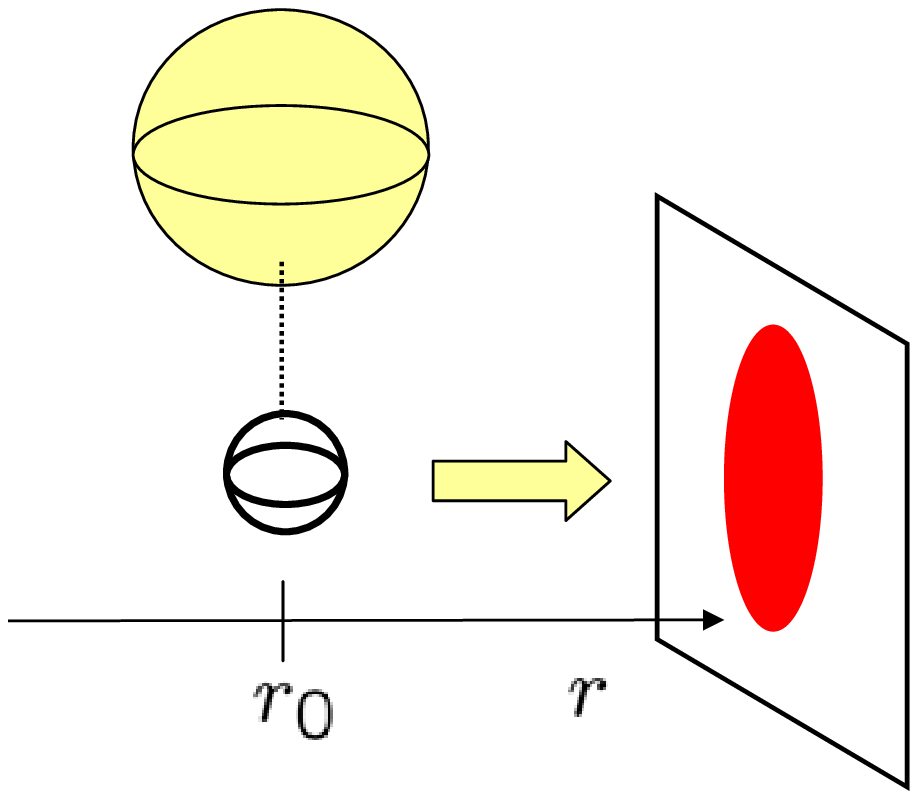}
\caption{The baryon vertices are distributed on a sphere. The arrow
denotes a projection onto the boundary space on the left.}
\label{fig3}
}
As we described, collective excitations on a nucleus can be
obtained by looking at a near horizon geometry of the D5 black-brane
solution, in the large $A$ limit. In general, computation of a fully
back-reacted solution in the type IIB supergravity is quite 
complicated, hence
we consider the following two simplifications. First, we need
to see only the near-horizon geometry of the D5-branes 
for our purpose, thus
we do not need the full geometry which asymptotes to the 
$AdS_5\times S^5$ geometry. Second, we may ignore the presence of the 
spikes (= fundamental strings attached to the D5-branes in
Fig.~\ref{fig2}) and the electric charges on the D5-branes, by the
following two reasons. The electric charge density induced on the
D5-brane worldvolume is proportional to $N/\lambda^{5/4}$, where
$N$ comes from the 5-form flux penetrating the $S^5$ as explained around
(\ref{css5}) and $\lambda^{5/4}$ is the volume of the $S^5$. So, 
in an appropriate limit $N, \lambda\to\infty$, this charge density can
be arbitrarily small. Furthermore, in Fig.~\ref{fig2}, the
left hand side of the $S^5$ of the D5-brane seems not really affected by
the fundamental strings which are attached to the right hand side of the
$S^5$ of the D5-brane. Hence we may just consider only the D5-branes
while ignoring the fundamental strings (and also the background flux
which generates the strings). 
This second simplification would be an assumption, as we will not show
in this paper that dynamics of the fundamental strings is irrelevant.

The first simplification can be realized by looking at the original
$AdS_5\times S^5$ metric around the position of the baryon vertex
$r=r_0$, 
\begin{eqnarray}
 ds^2 = \frac{r_0^2}{\alpha' \sqrt{2\lambda}} dx_{||}^2
+ \frac{\alpha' \sqrt{\lambda}}{r_0}dr^2 + dy^2 , 
\end{eqnarray}
where $x_{||}$ means $x^0, \cdots, x^3$ which are the spacetime
coordinates for the ${\cal N}=4$ SYM, and $dy^2$ is the metric on the
$S^5$. This large $S^5$ can be approximated by ${\bf R}^5$ for large
$N$. Resultantly, if we make the following rescaling of the coordinates,
\begin{eqnarray}
 \widetilde{x}_{||} \equiv \frac{r_0}
{\sqrt{\alpha' \sqrt{2\lambda}}} \; x_{||}
, \quad 
 \widetilde{r} \equiv \frac{\sqrt{\alpha' \sqrt{2\lambda}}}{r_0} \; r, 
\label{rescale}
\end{eqnarray}
the metric is nothing but just the flat spacetime 
metric in 10 dimensions. 
In these rescaled coordinates, we can easily obtain a supergravity
solution with a full back-reaction of the flat $A$ D5-branes
(if we ignore 
the effect of the fundamental strings as explained before
as the second simplification),
\begin{eqnarray}
 ds^2 & = & f^{-1/2}(-d\widetilde{t}^2 + dy^2) 
+ f^{1/2}(d\widetilde{x}_i^2
+ d\widetilde{r}^2), \\
e^{-2\phi}  & = & f.
\end{eqnarray}

\FIGURE[t]{
\includegraphics[width=3.5cm]{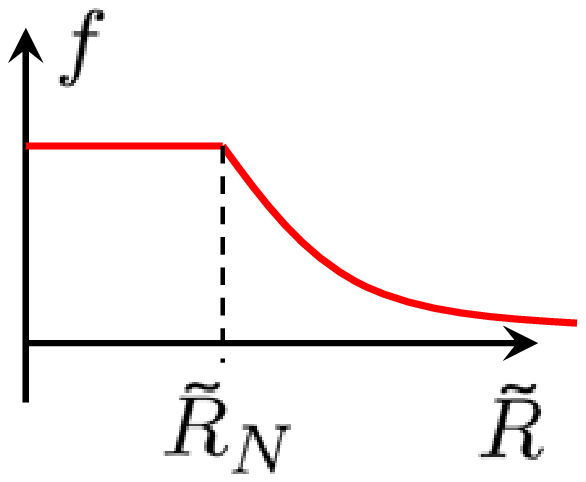}
\caption{The geometry of back-reacted D5-branes distributed on the $S^3$.}
\label{fig4}
}
For the case of $A$ D5-branes on top of each other, 
the harmonic function $f$ is given by
\begin{eqnarray}
 f = 1 + \frac{g_s \alpha' A}{\widetilde{R}^2}
\label{flarge}
\end{eqnarray}
where $\widetilde{R}\equiv \sqrt{\widetilde{x}_1^2 + \widetilde{x}_2^2+\widetilde{x}_3^2
+ \widetilde{r}^2}$ is the radial distance from the 
coincident D5-branes in the
rescaled coordinates. 

Our interest is the D5-branes distributed (smeared) uniformly on the
$S^3$. Let us suppose that the radius of the shell $S^3$ is
$\widetilde{R}_N$ in the rescaled coordinates. Then the function $f$ is
given by (\ref{flarge}) for $\widetilde{R}> \widetilde{R}_N$, while 
$f$ inside the shell is a constant, see Fig.~\ref{fig4}. 
The continuity
condition at the shell $\widetilde{R}=\widetilde{R}_N$ gives
\begin{eqnarray}
 f = 1 + \frac{g_s \alpha' A}{\widetilde{R}_N^2} \quad 
(\widetilde{R}< \widetilde{R}_N).
\label{fsmall}
\end{eqnarray}
The near horizon geometry with large $A$ is simply given by ignoring 
``$1+$'' in (\ref{flarge}) and (\ref{fsmall}).
This is the back-reacted near-horizon geometry which we need.

\section{Giant resonance of heavy nuclei}

According to the dictionary of the gauge/string duality, the
supergravity fluctuation spectrum in this geometry corresponds to the
spectrum of the theory on the D5-branes, that is, the spectrum of the
heavy nucleus. The simplest fluctuation among supergravity fields is the
dilaton fluctuation $\delta\phi$ whose equation of motion 
is\footnote{We assume that the fluctuation of the
dilaton in the Einstein frame and a fluctuation of the Ramond-Ramond
tensor fields 
are decoupled from each other.}  
\begin{eqnarray}
 \partial_M \left[ \sqrt{-\det g^{\rm (E)}}g^{{\rm (E)}MN} 
\partial_N \delta\phi
\right]=0.
\label{deltap}
\end{eqnarray}
Here note that the metric is in the Einstein frame, 
$g_{MN}^{\rm (E)} = e^{\phi/2} g_{MN}$.
The standard dictionary tells us that $\delta\phi$
corresponds partly to 
an operator ${\rm tr}[\Phi,\Phi]^2$ in the $U(A)$ SYM theory on the $A$
D5-branes. This operator describes a certain collective motion of the
constituent baryons, as the eigenvalues of the scalar field $\Phi$ on the
D5-branes denote the location of the D5-branes.

To solve this equation (\ref{deltap}) in our background,  we consider 
the form 
\begin{eqnarray}
 \delta \phi (\widetilde{x},\widetilde{r},y) 
= e^{i\widetilde{E}\widetilde{t}} 
g(\widetilde{R}), 
\end{eqnarray}
then (\ref{deltap}) reduces to an ordinary differential equation
\begin{eqnarray}
 \left[
\widetilde{R}^3 f(\widetilde{R})\widetilde{E}^2 
+ \partial_{\widetilde{R}} \widetilde{R}^3
\partial_{\widetilde{R}}
\right]g(\widetilde{R})=0.
\end{eqnarray}
With a new variable $s \equiv 1/\widetilde{R}^2$, this can be 
cast into a form of 
a Schr\"odinger equation, which can be solved. A simple 
dimensional analysis gives the energy
\begin{eqnarray}
 \widetilde{E}^2 = \frac{c}{g_s \alpha' A}
\label{c}
\end{eqnarray}
where $c$ is a dimensionless numerical constant.

Interestingly, the fluctuation spectrum in this background was already
computed in \cite{Kiritsis:2002xr}. It was shown that 
$c$ in (\ref{c}) is given by $c=1$, and 
it is merely a gap, above which the spectrum is continuous.
The motivation of 
\cite{Kiritsis:2002xr} was quite different in origin: the
$S^3$ spherical distribution 
was introduced there as a consistent UV cut-off for the gravity
dual of the 
D5/NS5 theory to probe little string theory. 

The mass gap (\ref{c}) with $c=1$ is written in the rescaled
coordinates, thus bringing it to the original coordinates, we find
\begin{eqnarray}
 E = \widetilde{E}\frac{r_0}{\sqrt{\alpha'\sqrt{2\lambda}}}
= \frac{(2\lambda)^{1/4}}{z_0} A^{-1/2}.
\label{gap}
\end{eqnarray}
Here in the last equality we have used (\ref{radius}).  
This expression (\ref{gap})
is the mass gap in the spectrum of the collective motion
of the scalar excitation on the supersymmetric heavy nucleus.
It should be called a monopole (spin zero) 
giant resonance of the 
supersymmetric nucleus.

In the following, we list 
key points of our result (\ref{gap}).
\begin{itemize}
\item
We find an interesting $A$ dependence in (\ref{gap}). 
The observed monopole 
giant resonances of heavy nuclei has $A^{-1/3}$ scaling, and our
$A$-dependence is different. However, we will see in the next section
     that a similar analysis in type IIA string theory 
gives $A^{-1/3}$.
We emphasize that our framework is capable of computing the $A$
dependence ---  it is quite important that 
we can compare it with the $A$ dependence observed in
nature. This is contrasted with the standard $N$ dependence of large $N$
QCD theories where real QCD is fixed to be at $N=3$. 
\item
The gap (\ref{gap}) is independent of $R_N$ but depends on $z_0$. 
$R_N$ is roughly the size of the nucleus, while $z_0$ is the size of
each baryon. In nature, both of these should be
determined by self-interactions of/among baryons, hence
the dependence found in (\ref{gap}) would be due to the 
supersymmetries. 
\item 
The gap (\ref{gap}) does not depend on $\alpha'$, due to (\ref{radius}) 
and the rescaling (\ref{rescale}).
All the physical quantities computed by the 
gauge/string duality should be independent of $\alpha'$, and our
      interpretation is consistent.
\end{itemize}

\section{Analysis in type IIA string theory} 

\FIGURE[h]{
\includegraphics[width=5cm]{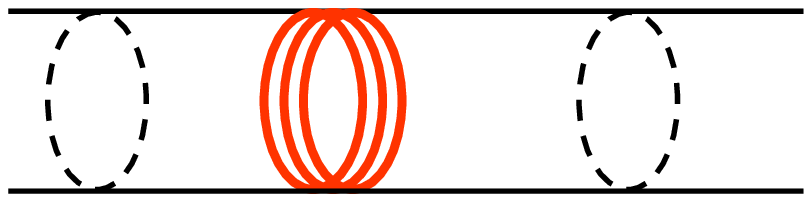}
\caption{$N$ D4-branes wrapping a circle.}
\label{fig5}
}
We can perform a similar analysis for type IIA brane configurations. 
To form a $SU(N)$ QCD-like gauge theory, 
let us consider $N$ D4-branes wrapping a circle, as in Fig.~\ref{fig5}. 
The radius of the circle is chosen to be $1/M_{\rm KK}$ so that the
Kaluza-Klein (KK) modes on the circle has the mass scale $M_{\rm KK}$. 
At low energy on the D4-branes, all the KK modes are massive and 
decoupled, then the theory there becomes the 3+1-dimensional 
${\cal N}=4$ SYM theory. The gauge coupling is given by the circle
compactification as 
\begin{eqnarray}
 \frac{1}{g_{\rm YM}^2} = \frac{2\pi}{M_{\rm KK}}
\frac{1}{(\pi)^2 g_s l_s}.
\end{eqnarray}
The gravity dual of this geometry is a near horizon geometry of
a BPS black 4-brane solution of the type IIA supergravity. 

Let us put baryon vertices which are D4-branes wrapping the $S^4$
of the geometry. 
As before, we specify the radial location of the baryon vertices as 
$r_0 = \alpha' U_0$ where $U_0$ is some physical scale, mimicking the
relation (\ref{radius}). This $U_0$ is a free parameter.
\FIGURE[t]{
\includegraphics[width=7cm]{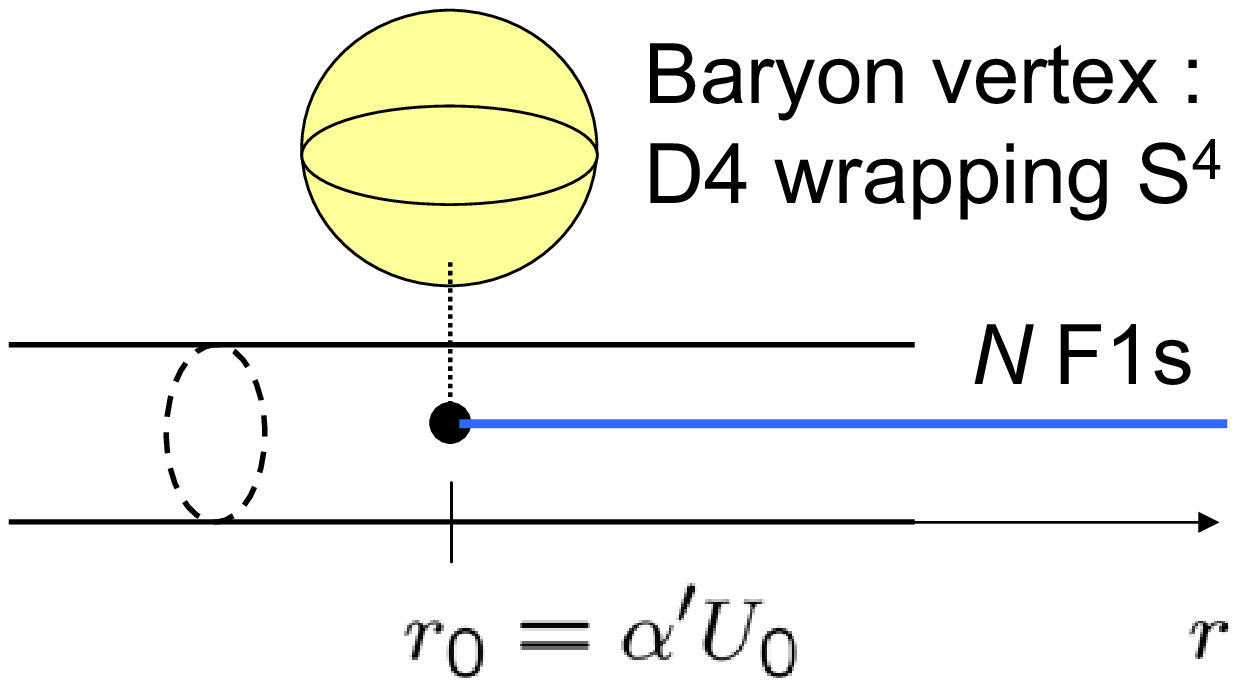}
\caption{The brane configuration in the type IIA string theory. }
\label{fig6}
}
Then, we distribute $A$ baryon vertices on a certain $S^4$ around 
$r=r_0$ in the space transverse to the $S^4$ of the background
geometry. This amounts to a procedure of having the ball-like
distribution of baryons in 3-dimensional space of the SYM theory, as one
can see by projecting the $S^4$ distribution of the D4-branes onto the
boundary space.

We consider a near horizon geometry of the $A$ D4-branes (baryon
vertices) and 
take a large $A$ limit. A similar analysis gives, this time, a discrete
spectrum of the dilaton fluctuation, 
\begin{eqnarray}
 \widetilde{E} = \frac{c}{3}
\sqrt{\frac{\widetilde{R}_N}{\pi l_s^3 \widetilde{g}_s A}},
\end{eqnarray}
with discrete values $c=2.23, 3.68, 4.75, \cdots$.
Here, again, $\widetilde{R}_N$ is the radius of the $S^4$ on which the
D4-branes are distributed, in the rescaled coordinates.
A similar rescaling back, $\widetilde{E} \to E$, 
$\widetilde{R}_N\to R_N$,  
$\widetilde{g}_s\to g_s$, results in
\begin{eqnarray}
 E = \left[\frac{8 M_{\rm KK}^3 U_0^3}{N}\right]^{1/4}
\frac13  R_N^{1/2} \; c \;  A^{-1/2}. 
\label{giant}
\end{eqnarray}
With the discrete values of $c$, we find a spectrum of a giant
resonance. 
Again, (\ref{giant}) does not depend on the string length 
$l_s^2 =\alpha'$, so it is a consistent observable in the gauge/string 
duality. 

Let us study the $A$ dependence.
If we substitute by hand $R_N\propto A^{1/3}$ which we adopt 
since it is expected from the nuclear density saturation (an observed
fact for realistic nuclei), we obtain
\begin{eqnarray}
 E \propto A^{-1/3}.
\end{eqnarray}
Interestingly, this dependence on $A$ is in accordance with the observed
monopole giant resonances.

\section{Summary and Discussions}

We have provided supersymmetric examples of a gravity dual of a heavy
nucleus. The dual is 
the near horizon limit of the back-reacted geometry of the baryon
vertices. The computation of the fluctuation spectra in this geometry
gives the holographic derivation of spectra of giant resonances of 
heavy nuclei. 

Generically, computations of supergravity backgrounds back-reacted 
due to the baryon vertices are 
complicated (see \cite{CHM} for the case of
spatially smeared baryons), but we find a simplification which 
leads to a known geometry of distributed black branes. There are a
lot of known geometries of this kind, and they would be of importance
for the application to nuclear physics, through the idea 
presented in this article and our
previous paper \cite{Hashimoto:2008jq},
 the holographic nuclei. On the
other hand, in the gauge/string duality, important back-reacted
geometries have been constructed (for example see 
\cite{Lunin:2007mj}), for example for large number of
fundamental strings in the bulk \cite{Yamaguchi:2006te}. It would be
interesting to explore possible relations between those geometries and
the holographic nuclei.

\acknowledgments 

I would like to thank H.~Y.~Chen, A.~Hashimoto, 
H.~Kawai, E.~Kiritsis, S.~Nakamura and H.~Ooguri for discussions. 
K.~H.~is partly supported by
the Japan Ministry of Education, Culture, Sports, Science and
Technology. The content of this paper was presented at RIKEN 
symposium 
``Towards New Developments in Field and String Theories'' on December
21st, 2008.


\begin{thebibliography}{99}

\bibitem{Maldacena:1997re}
  J.~M.~Maldacena,
  ``The large N limit of superconformal field theories and supergravity,''
  Adv.\ Theor.\ Math.\ Phys.\  {\bf 2}, 231 (1998)
  [Int.\ J.\ Theor.\ Phys.\  {\bf 38}, 1113 (1999)]
  [arXiv:hep-th/9711200].

\bibitem{Gubser:1998bc}
  S.~S.~Gubser, I.~R.~Klebanov and A.~M.~Polyakov,
  ``Gauge theory correlators from non-critical string theory,''
  Phys.\ Lett.\  B {\bf 428}, 105 (1998)
  [arXiv:hep-th/9802109].\\
  E.~Witten,
  ``Anti-de Sitter space and holography,''
  Adv.\ Theor.\ Math.\ Phys.\  {\bf 2}, 253 (1998)
  [arXiv:hep-th/9802150].\\
  O.~Aharony, S.~S.~Gubser, J.~M.~Maldacena, H.~Ooguri and Y.~Oz,
  ``Large N field theories, string theory and gravity,''
  Phys.\ Rept.\  {\bf 323}, 183 (2000)
  [arXiv:hep-th/9905111].


\bibitem{Gross-Ooguri}
  D.~J.~Gross and H.~Ooguri,
  ``Aspects of large N gauge theory dynamics as seen by string theory,''
  Phys.\ Rev.\  D {\bf 58}, 106002 (1998)
  [arXiv:hep-th/9805129].
\\
  E.~Witten,
  ``Baryons and branes in anti de Sitter space,''
  JHEP {\bf 9807}, 006 (1998)
  [arXiv:hep-th/9805112].

\bibitem{Hashimoto:2008jq}
  K.~Hashimoto,
  ``Holographic Nuclei,''
  Prog.\ Theor.\ Phys.\  {\bf 121}, 241 (2009)
  [arXiv:0809.3141 [hep-th]].

\bibitem{Imamura:1998gk}
  Y.~Imamura,
  ``Supersymmetries and BPS configurations on Anti-de Sitter
	space,''
  Nucl.\ Phys.\  B {\bf 537}, 184 (1999)
  [arXiv:hep-th/9807179].
\\
  C.~G.~.~Callan, A.~Guijosa and K.~G.~Savvidy,
  ``Baryons and string creation from the fivebrane worldvolume action,''
  Nucl.\ Phys.\  B {\bf 547}, 127 (1999)
  [arXiv:hep-th/9810092].


\bibitem{Callan:1997kz}
  C.~G.~Callan and J.~M.~Maldacena,
  ``Brane dynamics from the Born-Infeld action,''
  Nucl.\ Phys.\  B {\bf 513}, 198 (1998)
  [arXiv:hep-th/9708147].
\\
  G.~W.~Gibbons,
  ``Born-Infeld particles and Dirichlet p-branes,''
  Nucl.\ Phys.\  B {\bf 514}, 603 (1998)
  [arXiv:hep-th/9709027].

\bibitem{Balasubramanian:1998de}
  V.~Balasubramanian, P.~Kraus, A.~E.~Lawrence and S.~P.~Trivedi,
  ``Holographic probes of anti-de Sitter space-times,''
  Phys.\ Rev.\  D {\bf 59}, 104021 (1999)
  [arXiv:hep-th/9808017].

\bibitem{Kiritsis:2002xr}
  E.~Kiritsis, C.~Kounnas, P.~M.~Petropoulos and J.~Rizos,
  ``Five-brane configurations without a strong coupling regime,''
  Nucl.\ Phys.\  B {\bf 652}, 165 (2003)
  [arXiv:hep-th/0204201].
\\
  E.~Kiritsis, C.~Kounnas, P.~M.~Petropoulos and J.~Rizos,
  ``Five-brane configurations, conformal field theories and the
  strong-coupling problem,''
  arXiv:hep-th/0312300.

\bibitem{CHM}
  H.~Y.~Chen, K.~Hashimoto and S.~Matsuura,
  ``Towards a holographic model of color-flavor locking phase,''
  arXiv:0909.1296.

\bibitem{Lunin:2007mj}
  O.~Lunin,
  ``Strings ending on branes from supergravity,''
  JHEP {\bf 0709}, 093 (2007)
  [arXiv:0706.3396 [hep-th]].

\bibitem{Yamaguchi:2006te}
  S.~Yamaguchi,
  ``Bubbling geometries for half BPS Wilson lines,''
  Int.\ J.\ Mod.\ Phys.\  A {\bf 22}, 1353 (2007)
  [arXiv:hep-th/0601089].
\\
  O.~Lunin,
  ``On gravitational description of Wilson lines,''
  JHEP {\bf 0606}, 026 (2006)
  [arXiv:hep-th/0604133].



\end{thebibliography}
\end{document}